\documentclass[aip,jcp]{revtex4-1}
\pdfoutput=1
\usepackage{dcolumn}
\usepackage{graphicx}
\usepackage{graphicx,amssymb,amsmath}
\usepackage{graphics}
\usepackage{ulem}
\usepackage{color}
\def\r{\boldsymbol{r}}
\def\x{\boldsymbol{x}}

\def\kt{k_{\rm B} T}

\def\beq{\begin{equation}}
\def\eeq{\end{equation}}
\def\bea{\begin{eqnarray}}
\def\eea{\end{eqnarray}}

\begin{document}

\title{Electrostatics and aggregation: how charge can turn a crystal into a gel}
\author{Jeremy D.\ Schmit\footnote{ Present address: Department of Physics, Kansas State University {\tt schmit@phys.ksu.edu}}$^1$, Stephen Whitelam$^2$ and Ken Dill$^{1,3}$}
\affiliation{$^1$Department of Pharmaceutical Chemistry, University
of California,
 San Francisco, California 94158, USA \\
$^2$Molecular Foundry, Lawrence Berkeley National Laboratory, 1 Cyclotron Road, Berkeley, CA 94720, USA\\
$^3$Stony Brook University, Stony Brook, New York 11794, USA}

\begin{abstract}
The crystallization of proteins or colloids is often hindered
by the appearance of aggregates of low fractal dimension called
gels. Here we study the effect of electrostatics upon crystal and gel formation using an analytic model of hard spheres bearing point charges and short range attractive interactions. We
find that the chief electrostatic free energy cost of forming
assemblies comes from the entropic loss of counterions that render
assemblies charge-neutral. Because there exists more accessible
volume for these counterions around an open gel than a dense
crystal, there exists an electrostatic entropic driving force
favoring the gel over the crystal. This driving force increases with
increasing sphere charge, but can be counteracted by increasing
counterion concentration. We show that these effects cannot be fully
captured by pairwise-additive macroion interactions of the kind
often used in simulations, and we show where on the phase diagram to
go in order to suppress gel formation.
\end{abstract}

\maketitle
\section{Introduction}

Crystallizing proteins or colloids is difficult in part because the
growth of kinetic aggregates called gels can compete with crystal
nucleation~\cite{slabinski2007challenge,sear2007nucleation,noro1999role}.
Considerable insight into the competition between crystallization
and gelation has been obtained from models of spherical particles in
implicit solvent with pairwise-additive interactions,
modeling both the forces (hydrogen bonds, hydrophobic effects etc.)
that drive association, and the electrostatic
repulsions~\cite{gilson1988energetics} that act to suppress
association~\cite{kranendonk1988simulation,Asherie:1996,miller2004phase,
zaccarelli2005model,lu2008gelation}. Such models have shown how the
phase diagram varies with interaction strength and range, and have
revealed the presence of kinetically-arrested gel- and glass states
\cite{Dawson:2002}.  While pairwise-additive `screened' potentials
capture much of the important physics of electrostatics, they
neglect explicit counterion degrees of freedom. Here we show that
accounting for counterion entropies reveals that electrostatics does
more than simply prevent association: it can induce a thermodynamic
driving force that favors a gel over a crystal.

\section{Model} We consider a collection of hard spheres (macroions)
of radius $a$ in aqueous solution with added salt (we set $a=1.6$
nm, appropriate for a small protein). Spheres carry a point charge
$q$ at their centers, and interact with square-well attractions of
depth $-\epsilon$ and range $2a (\lambda -1)$. We are interested in
three phases: the solution phase of dissociated spheres, a
close-packed three-dimensional crystal, and a kinetically-stabilized
gel that we model as a one-dimensional rod of associated spheres. We
describe a crystallization window that is bounded on one side by the
thermodynamic stability of the crystal, and on the other by the
appearance of the kinetically favored gel.  We approximate the
kinetic threshold for gelation as the regime in which the chemical
potential of a linear aggregate drops below that of the solution;
since one-dimensional objects lack a nucleation barrier, we expect
gelation to preempt crystallization in this regime.  This threshold
reproduces qualitatively the gelation boundaries determined from
numerical and experimental studies of colloids and
proteins~\cite{noro1999role,Filobelo:2005,fu2003effect,Soga:1998,miller2004phase}
while allowing efficient computation of electrostatics.
\begin{figure*}[t]
\includegraphics[width=0.9\linewidth]{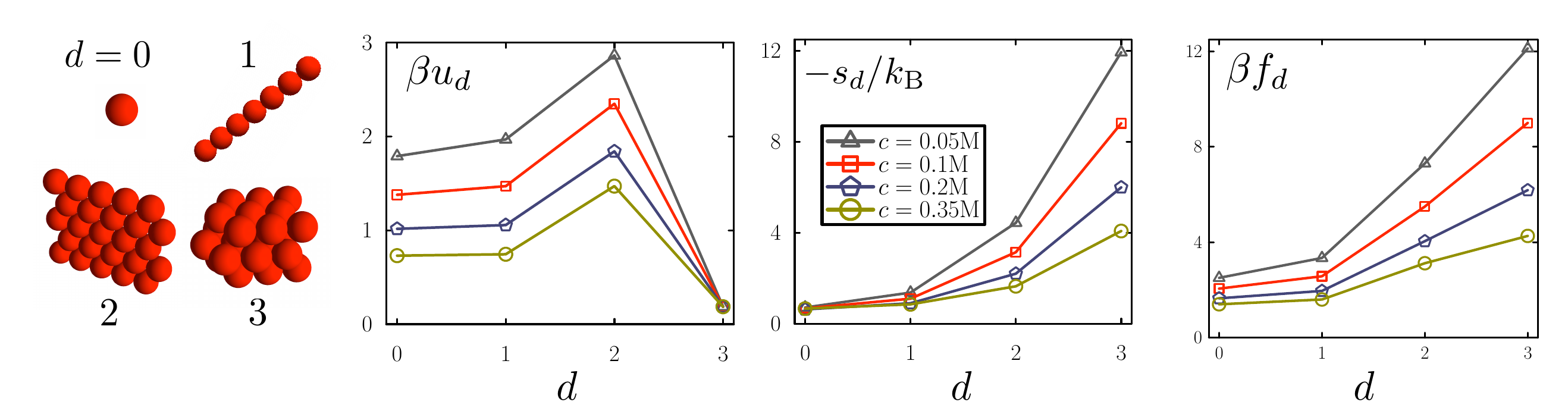}
\caption{\label{fig1} Counterion entropies dominate the electrostatic free energies of $d$-dimensional assemblies. We plot per-particle electrostatic energies, $u_d$, entropies, $s_d$, and free energies $f_d$ for each $d$-dimensional assembly (pictured left; the case $d=2$ is included for comparison only).  We have assumed a macroion charge $q=5e$ and have considered four different salt concentrations $c$. The electrostatic free energy cost for establishing each assembly is dominated by the entropy loss upon counterion confiment, and not by macroion-macroion repulsion energies.}
\end{figure*}

We write the chemical potentials of the zero-, one- and
three-dimensional phases (shown in Fig. \ref{fig1}) as
\begin{eqnarray}
\mu_0 = \ln \rho+\rho(4-3\rho)(1-\rho)^{-2} &\hspace{3 mm} & {\rm solution};\nonumber \\
\mu_1 = - \epsilon -\kt \ln(\lambda^3-1)  +\Delta f_1^{\rm es} & & {\rm gel};\label{eq:mu1}\\
\mu_3 = - z \epsilon/2  -3 \kt \ln(\lambda-1)+\Delta f_3^{\rm es} &
& \mathrm{crystal}. \nonumber
\end{eqnarray}
These expressions account for free energies of binding and
electrostatic free energies $\Delta f_d^{\rm es}$ associated with
each $d-$dimensional assembly ($d = 0$ refers to solution-phase
monomers, $d=1$ refers to chain-like aggregates and $d=3$ describes
the crystal phase). The chemical potential $\mu_0$ for the solution
phase contains the Carnahan-Starling expression for the free energy
of hard spheres as a function of $\rho$, the volume fraction of
spheres in solution~\cite{carnahan1969equation}. The expression
$\mu_1$ for the rod phase accounts for the fact that each particle
makes two pairwise contacts of energy $-\epsilon$. It also contains
an entropic term that accounts for the fact that a new particle can
add to the end of a rod anywhere within a shell of volume $\sim
(2a\lambda)^3-(2a)^3$ around the existing end particle. This
contribution accounts for both bending and vibrational entropies (we
neglect unimportant constant terms). We also neglect rod
branching~\cite{tlusty2000defect}, which at the level of
electrostatics amounts to assuming the characteristic linear extent
of a gel is long compared to the Debye length. The chemical
potential $\mu_3$ of the close-packed crystal accounts for the
energetic contacts between each particle and its $z$ nearest
neighbors (we set $z=12$), and the fact that each particle can
vibrate a distance of order $2a (\lambda-1)$ in each
direction~\cite{Asherie:1996}. We set $\lambda=1.2$ in our
calculations, motivated by the recognition that proteins typically
possess interactions of short
range~\cite{Asherie:1996,Prinsen:2006,Corezzi:2009,Young:2009}.

To these expressions we add the electrostatic free energies $\Delta f_d^{\rm es}$ of each
$d$-dimensional assembly. We calculated these free energies using the
Poisson-Boltzmann (PB) equation
\begin{equation}
\nabla_{\x}^2 \Phi_d=\sinh (\Phi_d), \label{pb}
\end{equation}
which we have written in terms of scaled spatial coordinates $\x
\equiv \kappa \r$ and a dimensionless potential $\Phi_d \equiv
e\psi_d/k_{\rm B}T$. $\kappa^{-1}$ is the Debye length, given by
$\kappa^2 \equiv 2 e^2 c/ \left(\epsilon k_{\rm B}T\right)$, where
$c$ is the concentration of ions in a reservoir of solvent in
osmotic equilibrium with the system. We assume our charged particles
to be in aqueous solution with positive and negative salt ions of
local concentration $c^\pm_d(\r) \equiv c e^{\mp \Phi_d(\r)}$. To
determine the potential $\Phi_d$ we solved Eq. (\ref{pb}) for given
assembly geometry. To simplify the solution of the PB equation we
approximated the rod as a smooth cylinder. For the crystal we used a
cell model~\cite{Alexander:1984,Prinsen:2006,Schmit:2010}. We have
also considered a two-dimensional assembly to gain insight into the
scaling behavior of the electrostatics, although we stress that one
would need to consider an {\em anisotropic} particle-particle
interaction to stabilize such a structure (see e.g.~\cite{tang2006self, whitelam2009impact}). The electrostatic
calculations are presented in detail in the appendix.

We then computed the electrostatic energy $U_d$ and entropy $S_d$ associated with each assembly.  The
energy $U_d$ stored in the electric field is
\begin{equation} U_d=\frac{\epsilon}{2}\int
d^3 \r \left( \nabla_{\r} \psi_d \right)^2, \label{field}
\end{equation}
where $\epsilon \equiv 80 \epsilon_0$ is the permittivity of water
($\epsilon_0$ is that of free space). The entropy $S_d$ quantifies
the cost of confining counterions and excluding coions from the
screening layer associated with a $d$-dimensional
assembly~\cite{oosawa1971polyelectrolytes}. The entropy cost per
unit volume of maintaining a screening layer of ion concentration
$c_{\rm s}$ in thermal contact with a reservoir of ions at
concentration $c$ is
\begin{eqnarray}
-S/k_{\rm B}&=&\int_{c}^{c_{\rm s}} {\rm d}c' \, \ln \left( c'/c \right) \nonumber \\
&=&c_{\rm s}\ln(c_{\rm s}/c)-c_{\rm s} + c.
 \label{ent}
\end{eqnarray}
The electrostatic free energy can then be written  $F_d = U_d - TS_d
= \int d^3 \r {\cal F}_d$, where the free energy density is
\begin{eqnarray}
{\cal F}_d&=&\frac{\epsilon}{2}\left( \nabla_{\r} \psi_d \right)^2
+k_{\rm B}T \left\{ c^+_d\ln(c^+_d/c) \right. \nonumber
\\&&\left. +c^-_d\ln(c^-_d/c)-c^+_d -c^-_d + 2 c
\right\}.
\label{free}
\end{eqnarray}
Note that the minimization of Eq. \ref{free} with respect to ionic
concentration results in Eq. \ref{pb} \cite{Netz:2000}.

Finally, to compute the electrostatic free energy cost of assembling
spheres into each $d-$dimensional structure, we computed the free
energy per particle in the assembly, $f_d=F_d/N$, and subtracted from
this the electrostatic free energy of the monomeric building blocks
in solution (determined by solving the PB equation for
an isolated sphere). The result is  $\Delta f_d^{\rm es}$.

\section{Results}

\subsection{Counterion entropy dominates the electrostatic free energy
of assembly.} Fig.~\ref{fig1} shows the electrostatic free energy
$f_d$ of a $d$-dimensional assembly, and its components, the
electrostatic energy $u_d$ and entropy $s_d$ (the case $d=2$ is
included for comparison, but it will not appear in our phase
diagram). We see that the energy per particle is non-monotonic in
aggregate dimensionality: the energy change upon condensation from
the free particle state to the crystalline state is {\em negative},
because the macroion-macroion repulsion is overcome by the
attraction between macroions and the counterions confined within the
crystal's cavities (assumed to be small compared to the Debye
length). By contrast, the electrostatic entropy cost increases
monotonically with aggregate dimensionality. Furthermore, these
entropies are much larger in magnitude than their energetic
counterparts, and so the electrostatic free energy cost becomes
larger as the structure increases in dimensionality. Although it
seems intuitively obvious that assembling charged spheres into a
crystal should meet with an electrostatic free energy cost, it is
not obvious, a priori, that the origin of this cost is entropic, and
not energetic~\cite{Warren:2002}.

Fig.~\ref{fig1} reveals that the electrostatic entropy alone is a
reasonable approximation of the electrostatic free energy cost of
assembly, a fact that has been long appreciated in polymeric systems
\cite{Warren:1997,Gottschalk:1998}. The behavior of the entropy with
$d$ can be understood from the following simple argument.
Electroneutrality requires that $q$ counterions be confined near
each macroion. Counterions are free to extend a distance
$\kappa^{-1}$ in each of the $3-d$ dimensions extending {\em away}
from the assembly, but are confined within a distance of order the
particle size $a$ in each of the $d$ dimensions {\em of} the
assembly. Thus the volume $v_{\rm localize}$ within which $q$
counterions must be localized scales as $v_{\rm localize} \sim a^d
\kappa^{d-3}$, and so the entropic cost per macroion of building the
screening layer scales as $-s_d/k \sim q\ln (c(a\kappa)^d/q\kappa^3)
$, or
\begin{equation}
-s_d/k \sim qd \ln (a \kappa).
\label{eq:scaling}
\end{equation}
This cost {\em increases} with $d$.

\subsection{Electrostatic interactions are not pairwise additive.} One
immediate consequence of the fact that counterion entropies dominate
the free energy cost for assembling charged particles is that
effective macroion interactions are {\em not} in general
pairwise-additive: the free energy difference $\Delta f_d^{\rm es}$
depends on the volume of space accessible to counterions, but does
not necessarily vary linearly with the number of contacts made by
particles in each assembly. We can define $\Delta_{31} \equiv z
\Delta f_1^{\rm es}$/2$\Delta f_3^{\rm es}$ as a measure of the
relative electrostatic cost of making a macroion-macroion contact in
a gel versus a crystal. When $\Delta_{31}$ is unity there is no
difference in this cost, the electrostatic effects can be adsorbed
into an effective pairwise interaction, such as a Yukawa
potential~\cite{hone1983phase}. However, when $\Delta_{31} \neq 1$
the electrostatic free energy is {\em non-additive}, rendered so by
counterion degrees of freedom.

\begin{figure}[ht]
\centering
\includegraphics[width=\linewidth]{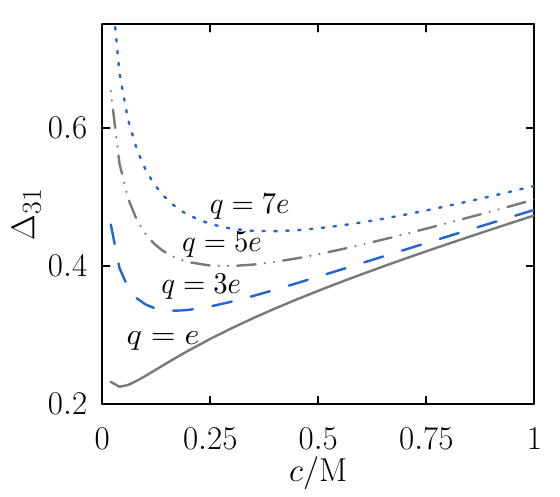}
\caption{\label{fig2} Relative electrostatic cost of making a macroion-macroion contact in a gel versus a crystal, $\Delta_{31}$, as a function
of salt concentration $c$ and macroion charge $q$. Only when $\Delta_{31}$ is unity is the electrostatics of assembly effectively pairwise-additive, which in general is not the case.}
\end{figure}

In Fig.~\ref{fig2} we plot $\Delta_{31}$ as a function of salt
concentration $c$ and particle charge $q$.  $\Delta_{31}$ is in
general less than unity, indicating that it is easier, per macroion
contact, to fit counterions around a gel than within a crystal.
$\Delta_{31}$  is also non-monotonic: in principle, at large salt
concentration strong electrostatic screening will ensure that
macroion sites are independent, with electrostatic free energies
becoming pairwise-additive (as seen in the plot, though, such
concentrations can be unattainably high). At low salt concentration
$\Delta_{31}$ again approaches unity.  This can be understood by
considering Eq. \ref{eq:scaling} for a cubic packing arrangement.
In this case $d$ is proportional to the number of nearest neighbors
and each additional pair of neighbors reduces the volume accessible
to the counterions by a factor $\kappa a$. This results in an
electrostatic free energy linear in the number of nearest neighbors.
However, at finite salt concentration the volume occluded by the
macroion (neglected in Eq. \ref{eq:scaling}) becomes a significant
perturbation to $v_{\rm localize}$, and the approximation of pairwise
additivity fails.

\subsection{Model phase diagram: how does one avoid gelation?} We turn
now to calculation of the model's phase diagram. We found the
solution-crystal coexistence line by setting $\mu_0=\mu_3$, and
estimated the nonequilibrium solution-gel coexistence line by
setting $\mu_0=\mu_1$. This line does not describe an equilibrium
coexistence: what we mean by it is the location at which there
exists a driving force for monomers to aggregate in a disordered
way. The crystal may be thermodynamically stable, but at moderate
degrees of supercooling or supersaturation it is separated from the
solution phase by a free energy barrier, and may require
considerable time to appear. By contrast, there exists no free
energy barrier to the formation of a linear aggregate, and so we
expect a gel to form readily below the nonequilibrium solution-gel
coexistence line.

\begin{figure}[ht]
\centering
\includegraphics[width=0.5\linewidth]{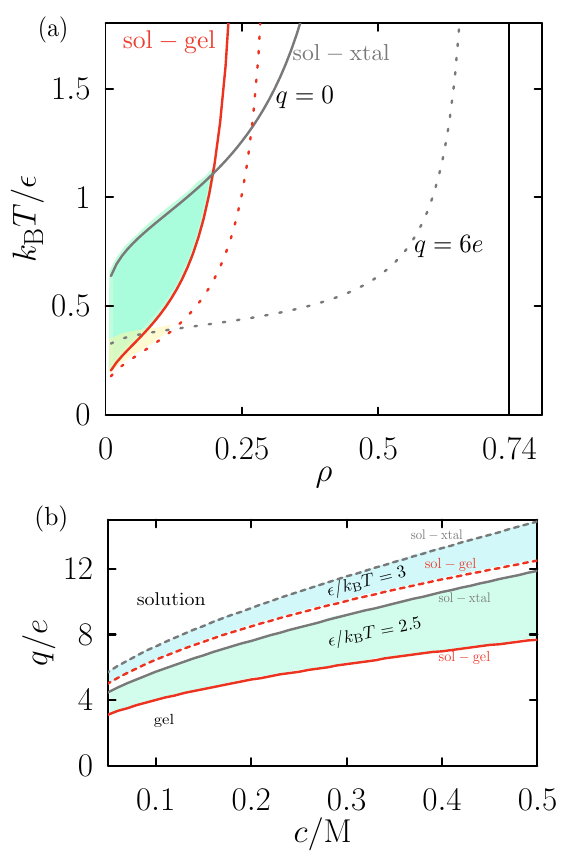}
\caption{\label{fig3} (a) Phase diagram in the temperature-density plane showing solution-gel (red) and crystal-solution (gray) coexistence lines, for macroion charges $q=0$ (solid) and $q=6e$ (dotted) for salt concentration 0.1 M. As macroion charge increases the crystal can be rendered stable by decreasing temperature, {\em but} counterion entropy increasingly favors the gel over the crystal. The `window' (shaded) within which the crystal is stable {\em and} the solution is stable against gelation therefore shrinks as charge increases. (b) Phase diagram in the charge-salt concentration plane for two bond strengths $\epsilon/k_{\rm B}T$. As charge increases, adding salt lessens the entropic difference between gel and crystal, widening the desired shaded region between the sol-gel and sol-crystal lines. Note that this region narrows as bond strength increases.}
\end{figure}

Our phase diagram in the temperature-density plane is shown in
Fig.~\ref{fig3}(a), for fixed salt concentration and for two
macroion charges. The close-packed crystal is stable below the gray
solution-crystal coexistence lines, but we expect a driving force
for gelation below the red solution-gel coexistence lines. This
behavior is similar to that seen in simulations of spheres with
short range attractions~\cite{fu2003effect}. The resulting `window'
of crystal stability and accessibility (shaded region between the
red and blue lines) is narrow, and {becomes narrower as macroion
charge increases}. This is a key finding of our study. As noted
before, the reason for this narrowing is the increasing cost of
confining counterions within a close-packed crystal relative to
around an open gel.

\subsection{High salt concentrations promote crystallization.} We have
seen that there are two consequences of increasing macroion charge:
the crystallization window narrows, and it shifts to smaller values
of $k_{\rm B}T/\epsilon$.  This latter effect is useful for protein
crystallization where we might not have the freedom to raise
temperature (lest it lead to denaturation): increasing protein
charge might be one way to move the crystallization window into the
region of $k_{\rm B}T/\epsilon$ accessible exprimentally.  The
tradeoff is that adding charge causes the window to narrow, but this
can be counteracted by adding salt. Fig.~\ref{fig3}(b) shows how,
for two fixed bond strengths $\epsilon/k_{\rm B}T$, one can alter
charge and salt concentration to widen the crystallization window.
Note that this window narrows as bond strength increases.

The shrinking of the crystallization window with the bond strength
can be understood by writing the net interaction between the
particles in terms of the attractive and repulsive components
$\epsilon_{\rm net}=-\epsilon + \epsilon_{\rm rep}$, where
$\epsilon_{\rm rep}=2\Delta f_3^{\rm es}/z$ is the electrostatic
{\em repulsion} energy per crystal contact.  In terms of these variables we can
express the solution-crystal coexistence line, given by
$\mu_0=\mu_3$, as
\begin{equation}
\epsilon_{\rm rep}=\epsilon+\frac{2}{z}(3k_{\rm B}T\ln(\lambda-1)+\mu_0).
\label{eq:SPDxtal}
\end{equation}
Similarly, we can express the solution-gel coexistence line as
\begin{equation}
\epsilon_{\rm rep} = \frac{\epsilon}{\Delta_{31}} +
\frac{1}{\Delta_{31}}(k_{\rm B}T\ln(\lambda^3-1) + \mu_0),
\label{eq:SPDgel}
\end{equation}
where we have used $\Delta_{31}$ to write the electrostatic free
energy of the gel in terms of $\epsilon_{\rm rep}$.

\begin{figure}[ht]
\centering
\includegraphics[width=\linewidth]{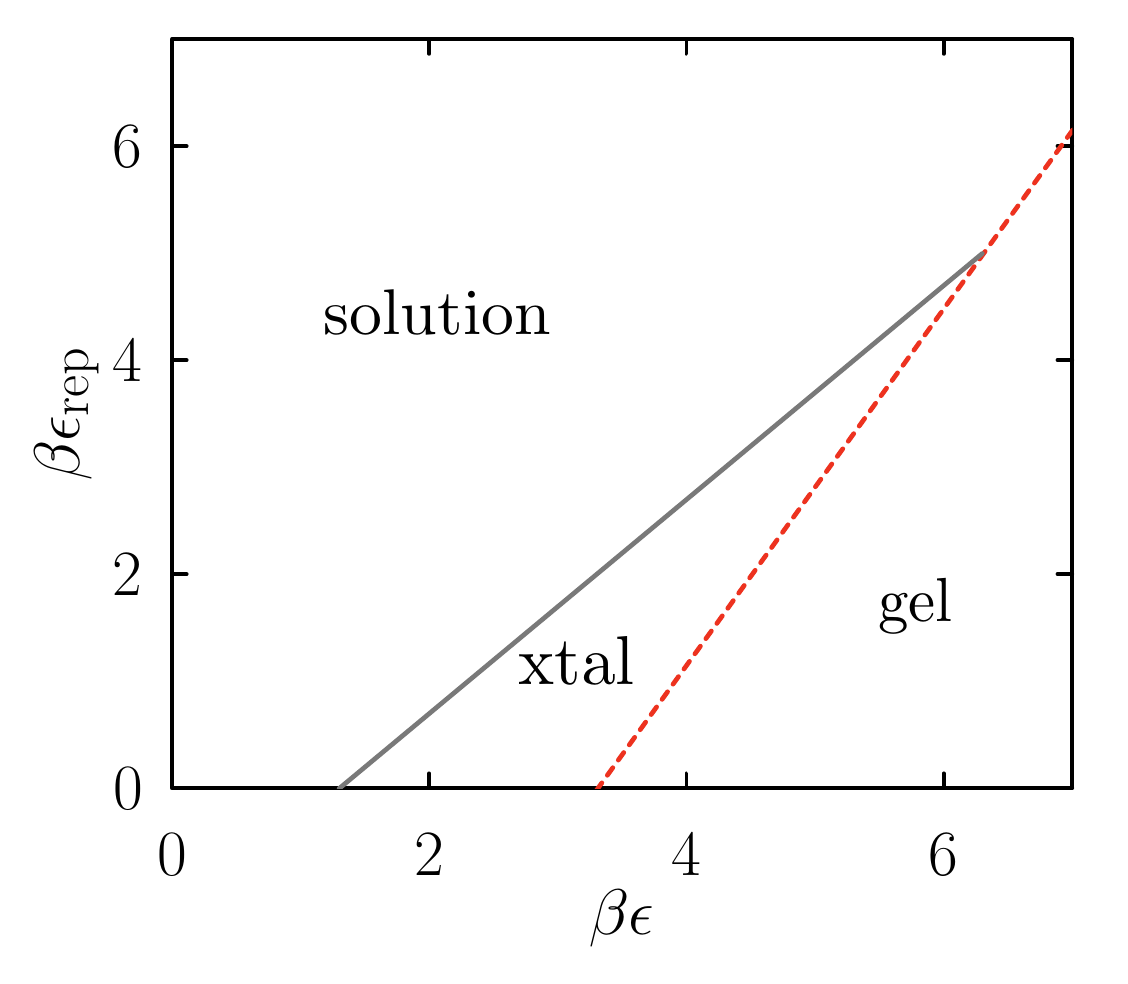}
\caption{\label{fig:SPD} Phase diagram showing the lines
$\mu_0=\mu_1$ (dotted) and $\mu_0=\mu_3$ (solid) as a function of the
attractive and repulsive contributions to the net interaction (here $\Delta_{31}=0.6$ and $c=0.04$).}
\end{figure}

Eqs. \ref{eq:SPDxtal} and \ref{eq:SPDgel} are plotted in Fig.
\ref{fig:SPD}.  We see that strongly attractive particles (large
$\epsilon$) can be brought within the crystallization window by
increasing $\epsilon_{\rm rep}$.  However, this window becomes
narrower with increasing $\epsilon$, and above a critical attraction
vanishes entirely. This happens because strongly attractive particles
require large charges to weaken particle-particle interactions enough to destabilize the gel, but
large charges in turn recruit dense screening layers which act to favor the gel over the crystal.  This effect can be mitigated by working under
conditions for which electrostatic interactions are approximately pairwise-additive (large
$\Delta_{31}$).  Increasing the value of $\Delta_{31}$ decreases the
slope of the solution-gel coexistence line (dotted line in Fig.
\ref{fig:SPD}). This widens the crystallization window and
increases the critical value of $\epsilon$ at which the solution-gel
line overtakes the solution-crystal line.  In practice,
$\Delta_{31}$ can be increased by simultaneously increasing salt
concentration and particle charge.  For example, within our model
system a macroion charge $q=2.8$ in 100 mM salt, or a charge of $q=8.1$ in 800 mM salt, both destabilize crystal
bonds by $0.5 \, k_{\rm B}T$. However, the
former condition destabilizes bonds in the gel phase by only $0.17 \,
k_{\rm B}T$, while the latter condition destabilizes bonds by $0.25 \,  k_{\rm B}T$.  Thus,
high salt/high charge conditions are more conducive to
crystallization than low salt/low charge conditions.  We speculate
that the addition of multi-valent counterions may further reduce the entropic cost of neutralizing
the crystal, athough a quantitative analysis of this scenario is beyond the
limitations of our PB model.

\section{Conclusion}

We have studied a minimal model of the crystallization and gelation
of charged attractive spheres in aqueous salt solution. We find that
there exists an electrostatic driving force that favors the gel over
the crystal, because there exists more ion-accessible volume for
confined counterions around a gel than within a crystal. This effect
renders the effective interactions between macroions
non-pairwise-additive, and is responsible for a narrowing of the
crystallization window as macroion charge is increased.

{\em Acknowledgements.} This work was performed as part of a User project at the Molecular Foundry, Lawrence Berkeley
National Laboratory, which is supported by the Office of Science,
Office of Basic Energy Sciences, U.S. Department of Energy, under
Contract No. DE-AC02-05CH11231. K.D. appreciates the support of NIH
grant GM34993, Defense Threat Reduction Agency grant IACRO-B0845281,
and the support of the Sandler Family Foundation.  The authors would
like to thank Fyl Pincus and Martin Muschol for critical readings of
the manuscript.

\appendix

\section{Poisson-Boltzmann free energies}

For the environmental conditions considered in this paper we find
that solutions $\Phi_d$ of the Poisson-Boltzmann (PB) equation [Eq.
(2) in the main text] are closely approximated by solutions
$\Phi^{\rm lin}_d$ of its linearized counterpart, the Debye-Huckel
(DH) equation. We present numerical results derived from the PB
equation, but also present, for the reader's convenience, the
well-known analytic solutions of the DH equation. For completeness,
we also consider a 2$d$ membrane (in 2d the PB equation has an
analytic solution). We draw the same qualitative conclusions from
both levels of theory.

To compute the electrostatic free energy cost of assembling the
building blocks into each $d$-dimensional structure, we compute the
free energy per particle in the assembly, $f_d=F_d/N$ and subtract
from this the electrostatic free energy of the monomeric building
blocks in solution. The per-particle electrostatic free energy of
assembly is then $\Delta f_d=f_d-f_0$.  To determine $f_0$ we solve
the PB equation for the potential of an isolated sphere of radius
$a$ carrying charge $q$. The corresponding solution of the DH
equation is
\begin{equation}
\Phi_0^{\rm lin}(x)=\frac{\kappa \ell_{\rm B} e^{\kappa a}}{1+\kappa
a}\frac{e^{-x}}{x}.
\end{equation}
Here $\ell_{\rm B} \equiv e^2/\left(4 \pi \epsilon k_{\rm
B}T\right)$ is the Bjerrum length.

{\em 1D assembly.}  We treat the 1D assembly as an infinitely long
cylinder of radius $a$ and linear charge density $\rho=q/2a$, and
obtain $\Phi_1$ from the PB equation in plane polar coordinates. The
corresponding solution of the DH equation
is~\cite{andelman2006introduction}
\begin{equation}
\Phi_1^{\rm lin}(x) =  \frac{e}{\kt} \frac{q}{4 \pi a^2 \epsilon
\kappa {\rm K}_1 (\kappa a)} {\rm K}_0(x).
\end{equation}
Here $x$ is the scaled distance from the cylinder center and ${\rm
K}_n$ is the $n^{\rm th}$ order modified Bessel function of the
second kind.

{\em 2D assembly.}  We treat each surface of the 2D assembly as an
infinitely extended plane carrying areal charge density $\sigma=q
{\cal A}_2/2 \pi a^2$. Here ${\cal A}_2\simeq 0.91$ is the area
occupied by close-packed spheres in a sheet. The exact solution of
the PB equation in planar geometry at a scaled distance $x$ from the
plane surface is
\begin{equation}
\Phi_2(x)=2\ln
\left(\frac{1+e^{-x}\tanh(\phi_0/4)}{1-e^{-x}\tanh(\phi_0/4)}\right),
\label{eq:planepot}
\end{equation}
where $\phi_0 \equiv 2\sinh^{-1}\left[2\pi \ell_{\rm B}
\sigma/(\kappa e)\right]$ is the electrostatic potential at the
plane surface. From Eqs. (3) and (4) in the main text we find
\begin{equation}
U_2=A_{\rm sheet} \frac{\kappa k_{\rm B}T }{\pi \ell_{\rm
B}}\frac{\lambda^2}{1-\lambda^2},
\end{equation}
and
\begin{equation}
S_2/k_{\rm B}=A_{\rm sheet}\frac{8
c_0}{\kappa}\frac{3\lambda^2-2\lambda\ln\left(\frac{1+\lambda}{1-\lambda}\right)}{1-\lambda^2},
\end{equation}
where $\lambda \equiv \tanh(\phi_0/4)$ and $A_{\rm sheet}$ is the
area of the sheet.

{\em 3D assembly.}  Following previous work on colloidal
systems~\cite{Alexander:1984,Prinsen:2006,Schmit:2010} we treat the
3D assembly as a collection of spherical macroions of radius $a$,
each of which is surrounded by a spherical aqueous cavity of radius
$b$. We assume that macroions are close-packed at volume fraction
${\cal A}_3 = (a/b)^3\simeq 0.74$. To calculate the electrostatic
free energy of this assembly we assume that the electric field on
the surface of each aqueous cavity vanishes~\cite{Schmit:2010}
(valid for assemblies whose characteristic linear size is much
greater than the Debye length). We also assume that the field at the
macroion surface is unaffected by the presence of salt and
counterions. We find the potential $\Phi_3$ from the appropriate PB
equation. This solution is closely approximated by linearizing the
PB equation around the average potential between macroions. By
writing $\Phi_3(x) = \Phi_3^{\rm lin}(x)+\bar{\phi}$, where $x$ is
the scaled distance from the macroion center, and the mean potential
\begin{equation}
\bar{\phi} \equiv \sinh^{-1}\left(\frac{3 q{\cal A}_3}{8 \pi c_0 a^3
(1-{\cal A}_3)}\right) \label{eq:meanpotential}
\end{equation}
is given by a jellium model~\cite{Warren:2002,Schmit:2010}, we
impose the boundary conditions described above and find, after some
algebra,
\begin{eqnarray}
\Phi_3^{\rm lin}(y)&=&\frac{\alpha^2
E_0}{e^{-\alpha+\beta}(\alpha+1)(\beta-1)-e^{\alpha-\beta}(\alpha-1)(\beta+1)}\times \nonumber \\
&&\left(\frac{e^{\beta-y}(\beta-1)}{y}+\frac{e^{y-\beta}(\beta+1)}{y}\right)
- \tanh \bar{\phi}.
\end{eqnarray}
Here $E_0 \equiv qe\kappa (\cosh \bar{\phi})^{1/2}/\left(4 \pi
\epsilon k_{\rm B}T \right)$; $y \equiv (\cosh
\bar{\phi})^{1/2}\kappa r$; $\alpha \equiv (\cosh
\bar{\phi})^{1/2}\kappa a$; and $\beta \equiv (\cosh
\bar{\phi})^{1/2}\kappa b$.

Our model is likely to become unreliable at high salt concentrations
and at low values of the packing fractions ${\cal A}_{2,3}$. Under
such conditions the approximation of taking filament and sheet
surfaces to be smooth becomes unrealistic because the aqueous
volume of surface corrugations becomes comparable to the total
volume of the screening layer (accounting for steric corrections to
the PB equation~\cite{borukhov1997steric} then becomes necessary).
In order to neglect these corrugations, it is necessary that both
the Debye length and the Gouy-Chapman length, $\ell_{\rm GC}=e/2\pi
\ell_B \sigma$, exceed the characteristic length scale of the
surface cavities. For this reason we restrict our analysis below to
salt concentrations below $c = 0.5$ M and $q=10$, for our chosen
macroion radius $a=1.6$ nm.

\end{document}